\documentclass[a4paper,11pt]{article}
\usepackage{pos}
\usepackage{wrapfig}
\usepackage{lineno}

\newcommand{\nc}{\newcommand}
\nc{\denselist}{\setlength{\itemsep}{0cm} \setlength{\parskip}{0cm}}
\nc{\ev}{\mathrm{eV}}
\nc{\mev}{\mathrm{MeV}}
\nc{\gev}{\mathrm{GeV}}
\nc{\kev}{\mathrm{keV}}
\nc{\tev}{\mathrm{TeV}}
\nc{\pev}{\mathrm{PeV}}
\nc{\eev}{\mathrm{EeV}}
\nc{\zev}{\mathrm{ZeV}}
\nc{\tv}{\mathrm{TV}}

\title{HAWC measurements of the energy spectra of cosmic ray protons, helium and heavy nuclei in the TeV range}

\ShortTitle{Cosmic ray composition studies with HAWC}

\manuallySeparateAuthors
\author*[a]{J.C. Arteaga-Velazquez} 
\author{for the HAWC Collaboration} 

\affiliation[a]{Instituto de Física y Matemáticas, Universidad Michoacana, \\ Morelia, Mexico}

\emailAdd{juan.arteaga@umich.mx}

\abstract{Current knowledge of the relative abundances and the energy spectra of the elemental mass groups of cosmic rays in the $10 \, \tev - 1 \, \pev$ interval is uncertain. This situation prevents carrying out precision tests that may lead to distinguish among the existing hypotheses on the origin and propagation of $\tev$ cosmic rays in the galaxy. In order to learn more about the mass composition of these particles, we have employed HAWC data from hadron induced air showers in order to determine the spectra of three mass groups of cosmic rays: protons, helium and heavy nuclei with $Z > 2$. The energy spectra were estimated by using the Gold unfolding technique on the 2D distribution of the lateral shower age against the estimated primary energy of events with arrival zenith angles smaller than 45 degrees. The study was carried out based on simulations using the QGSJET-II-04 model. Results are presented for primary cosmic-ray energies from $10 \, \tev$ to $251 \, \tev$. They reveal that the aforementioned cosmic ray spectra exhibit fine structures within the above primary energy range.}

\FullConference{37$^{\rm{th}}$ International Cosmic Ray Conference (ICRC 2021)\\
		July 12th -- 23rd, 2021\\
		Online -- Berlin, Germany}

\begin{document}
\maketitle

\section{Introduction}

 One of the challenges for cosmic ray research in the $\tev$ regime is the measurement of the spectrum and the relative abundances of the different elemental species. This knowledge is important if one wants to achieve  a complete understanding of  cosmic ray physics at $\tev$ energies and to obtain a comprehensive picture on the origin of cosmic rays from  $\tev$ to $\pev$ energies. In this regard, the HAWC observatory could help to probe the composition and energy spectrum of $\tev$ cosmic rays. 
 
 HAWC is a high altitude detector array for measuring  extensive air showers (EAS) at $\tev$ energies of hadronic and gamma ray origin \cite{Hawc17, Hawc17a}.  It is located at $4100 \, \mbox{m}$  a.s.l on a terrace on the Sierra Negra Volcano in the central east part of Mexico. The main part of the instrument consists of a compact array of $300$ water Cherenkov detectors distributed over a $22,000  \, \mbox{m}^2$ surface. The cylindrical detectors are $7.3 \, \mbox{m} \, \mbox{diameter} \times 4.5 \, \mbox{m}  \, \mbox{deep}$ and each contains $4$ photomultipliers (PMT). These units  provide data on the local charge and arrival times of the EAS front. The shower reconstruction in HAWC allows to find the impact point of the shower core, the arrival direction of the cascade and the lateral distribution of the measured charge per PMT (LDF) of the event. The primary energy of the shower is calibrated using a maximum likelihood estimation \cite{Hawc17} that involves a comparison between the measured LDF with QGSJET-II-04 \cite{qgsjetii4} predictions for proton primaries  . The lateral shower age per event is also obtained and it is estimated by fitting the LDF data with an NKG function applying the minimum $\chi^2$ method \cite{Hawccrab19}.
 In this work, we have performed an unfolding analysis of the 2D dimensional distribution of shower age vs primary energy for selected HAWC data to estimate the energy spectra of H, He and heavy nuclei ($Z > 2$). We have used Monte Carlo (MC) simulations with the QGSJET-II-04 model to derive the response matrices employed in the unfolding analysis. The procedure and results will be described in the next sections. \vspace{-0.8pc}
  

\section{Analysis procedure}

 We have selected a data set of $5.17 \times 10^{10}$ EAS collected during an effective time of $T_{\mbox{eff}} = 3.21 \, \mbox{years}$. For the analysis, we only used data that passed the shower core and arrival direction reconstruction procedures, with zenith angles $\theta$ below $45^\circ$, a fraction of hit PMTs $\geq 0.2$, number of activated PMTs in a radius of $40 \, \mbox{m}$ around the shower core equal to $40$, shower age $s = [1.0, 3.2]$, and primary energy $\log_{10}(E/\gev) = [3.5, 6.5]$. These selection criteria help to reduce the systematic uncertainties on the reconstructed data and  the unfolded results. With these cuts,  above $10 \, \tev$, the shower core resolution is smaller than $17 \, \mbox{m}$ and the angular resolution is not larger than $0.6^\circ$. On the other hand, the energy resolution $\sigma \log_{10}(E/\gev)$ is $< 0.35$.

\begin{figure}[!t]
     \centering
     \footnotesize
     \includegraphics[width=75mm]{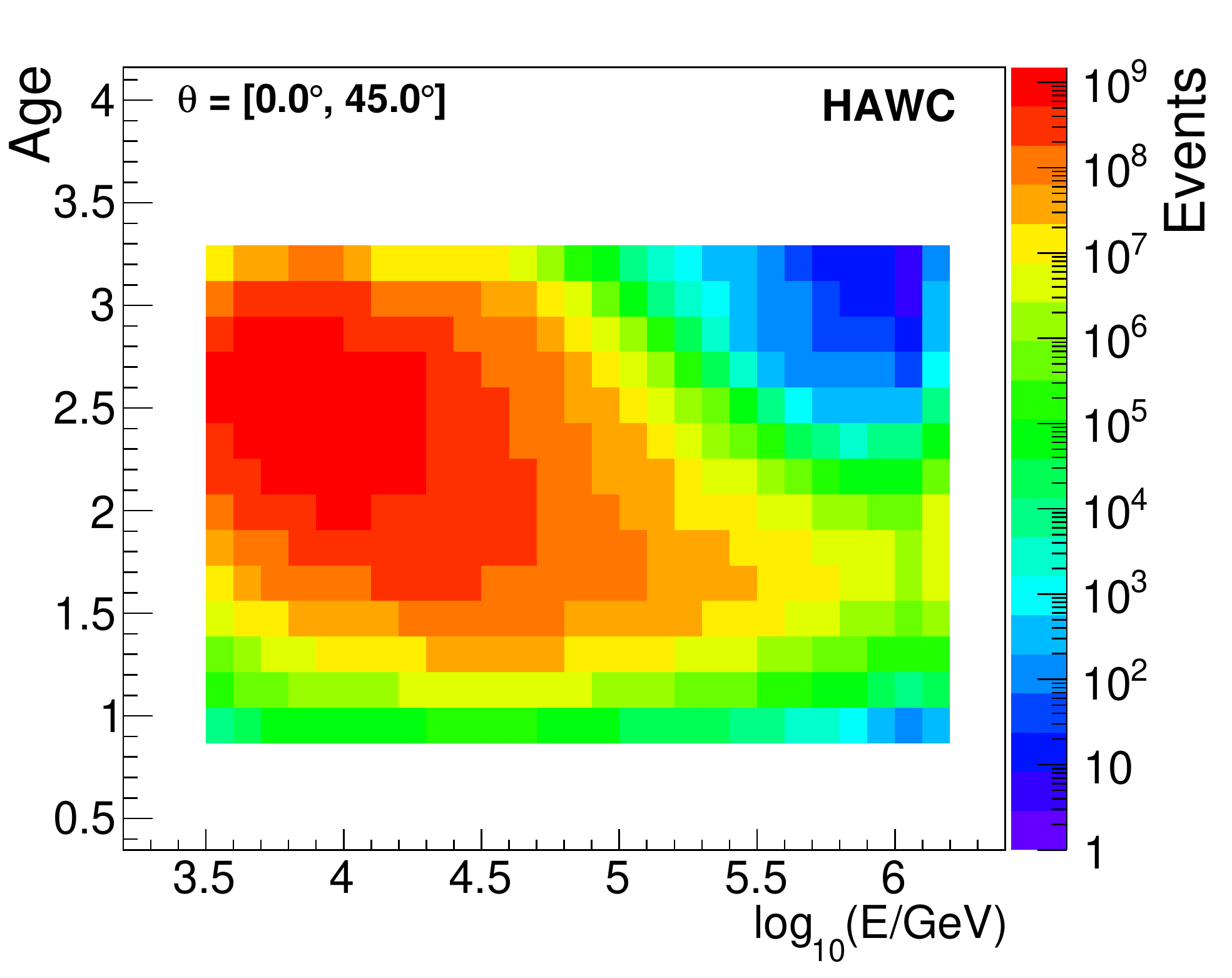}
     \includegraphics[width=75mm]{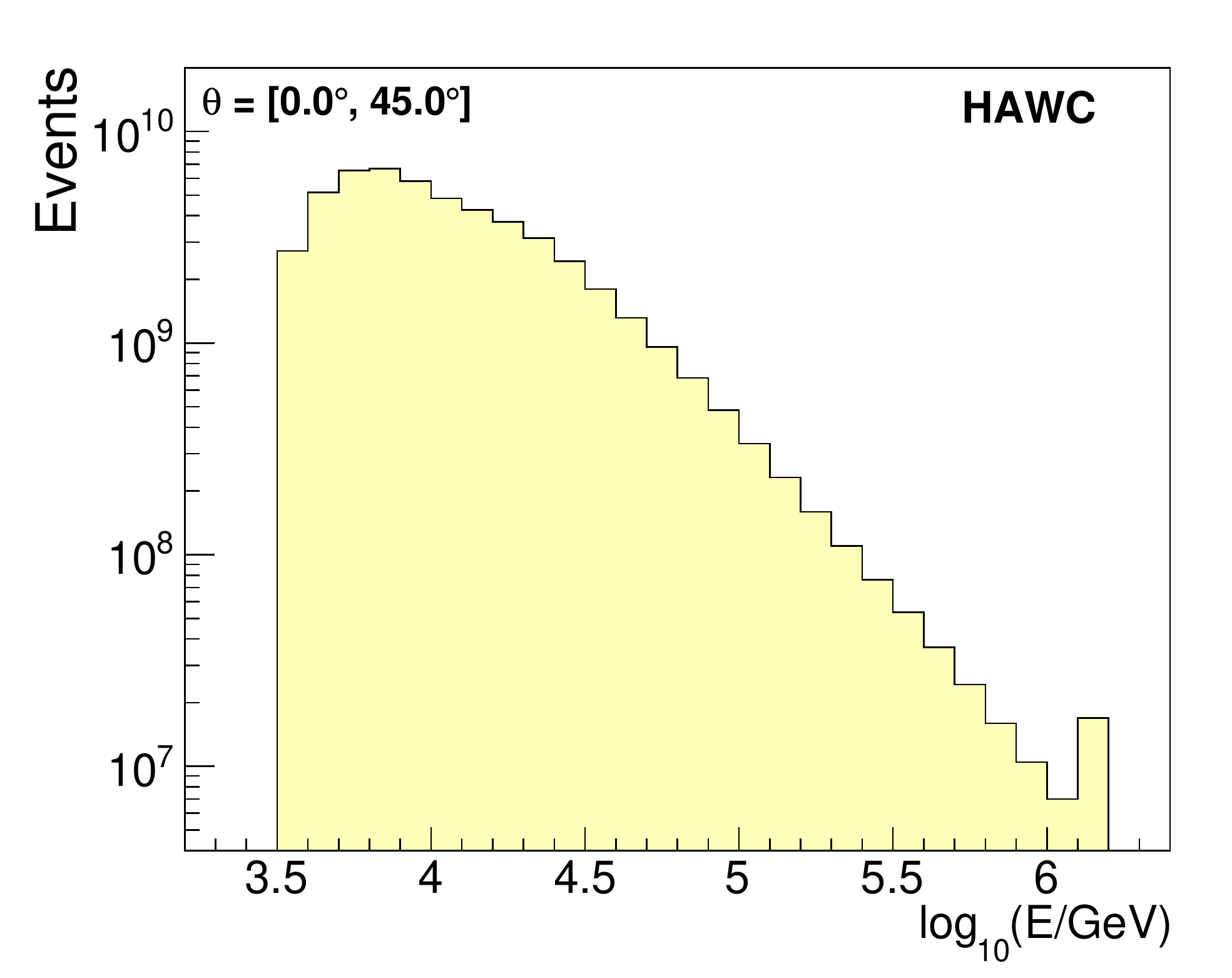} 
     \caption{ 
     \textit{Left}: 2D distribution of the measured shower age vs the reconstructed primary energy for the selected HAWC data. 
     \textit{Right}: Measured HAWC energy histogram after selection cuts. 
    }
  	\label{data}
  	\vspace{-1pc}
\end{figure}

\begin{wrapfigure}[15]{r}{0.5\textwidth}
  \vspace{-2pc}
  \begin{center}
     \footnotesize
    \includegraphics[width=75mm]{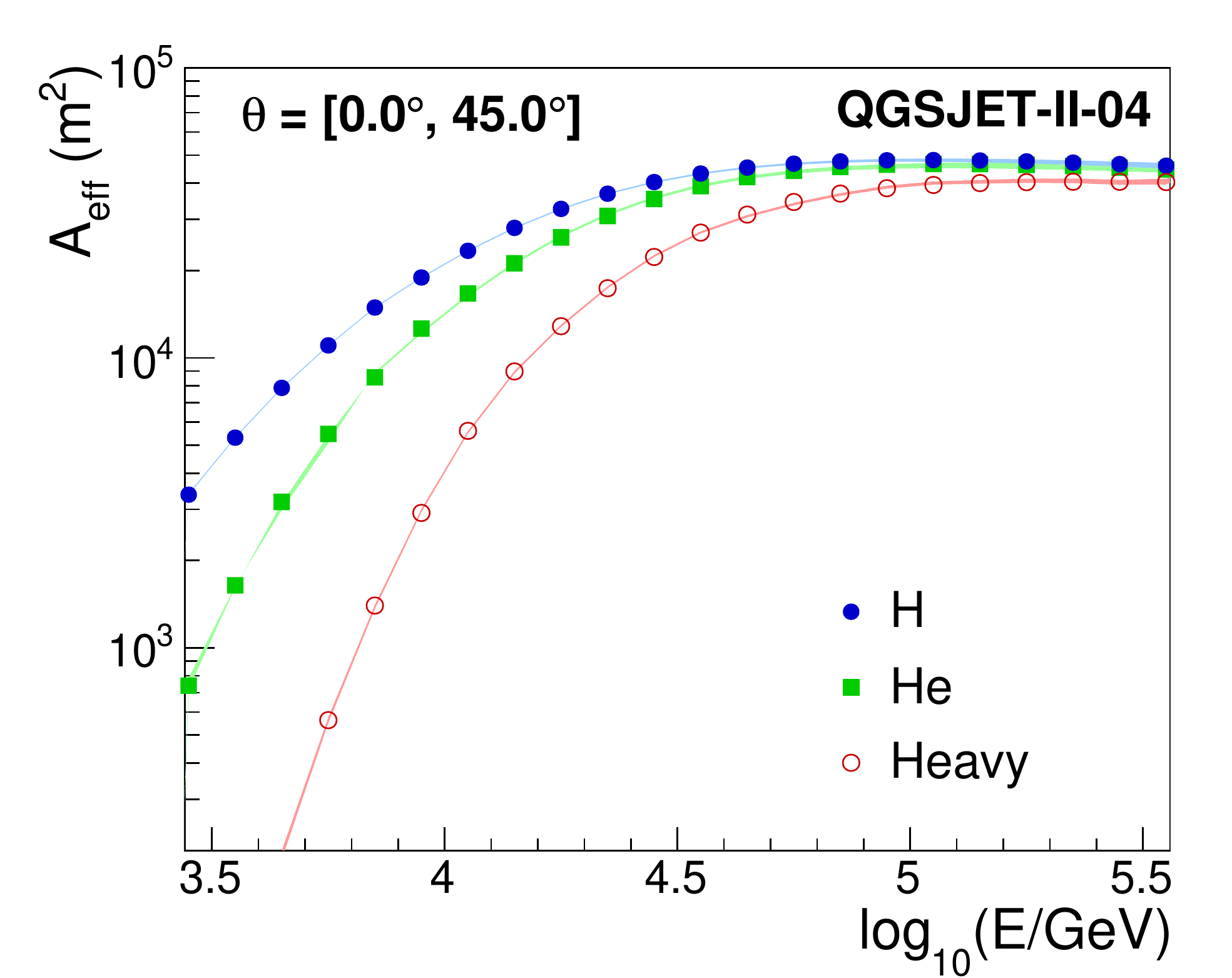}
  \end{center}
     \caption{HAWC's effective area for different mass groups obtained with MC simulations.}
  	\label{eff}
\end{wrapfigure}

 The measured age vs energy distribution, $n(s,\log_{10}E)$, and the primary energy histogram of the data are shown in fig.~\ref{data}. We employed bins of size $\Delta s = 0.17$ and  $\Delta \log_{10}(E/\gev) = 0.1$ to build the histograms. The measured 2D distribution is related to  the primary spectra of the elemental species of cosmic rays, $\Phi_{j}(E_{T})$, trough the following equation:
 \begin{equation}
     n(s,\log_{10}E) = T_{\mbox{eff}} \,  \Delta \Omega \, \sum_{j = 1} \sum_{E_{T}} P_j(s, \log_{10}E| \log_{10}E_{T}) \, A_{\mbox{eff}, j}(E_{T})\, \Phi_{j}(E_{T}) \, \Delta E_{T},
     \label{eq1}
 \end{equation}
 where $E_{T}$ is the true primary energy, $j$ runs over each primary nucleus that contributes to the measured distribution,  $\Delta \Omega$ is the solid angle interval, $A_{\mbox{eff}, j}(E_{T})$ is the effective area for the $j$ primary nucleus and $P_j(s, \log_{10}E| \log_{10}E_{T})$ is a response matrix associated with the probability that a given cosmic ray element $j$ with true energy $E_T$, which is detected at HAWC, creates an EAS that is reconstructed with a primary energy $E$ and shower age $s$. For this analysis, $j$ runs only over three mass groups: H, He and heavy nuclei ($Z > 2$). The effective areas and the response matrices are obtained using MC simulations. The MC data set includes $4.9 \times 10^{10}$ EAS simulated with CORSIKA \cite{Heck:1998vt} and the hadronic interaction models QGSJET-II-04 and FLUKA \cite{Fluka} for eight primary nuclei: H, He, C, O, Ne, Mg, Si and Fe, with $\theta < 65^\circ$. The simulations were produced with $E^{-2}$ primary energy spectra from $5 \, \gev$ to $10 \, \pev$ (energy per particle), but they were appropriately reweighted to reproduce the composition model for cosmic rays of \cite{Hawc17}, which was obtained from fits to PAMELA \cite{pamela}, AMS-2 \cite{ams14, ams15} and CREAM I-II \cite{cream09, cream11} data. The effective areas for the H, He and heavy mass groups of cosmic rays are shown in fig.~\ref{eff}. They were calculated using the procedure described in \cite{Hawc17} and our MC data.
 
  In order to find the energy spectra from eq.~(\ref{eq1}), we have applied the Gold's unfolding \cite{Gold64} procedure as described in \cite{Ulrich01}. The results were checked with the reduced cross-entropy unfolding method \cite{Crossentrop}. The priors used in the unfolding technique are the spectra from our nominal composition model with appropriate normalization factors for the spectrum of each mass group, which allow to minimize the $\chi^2$ differences between the observed 2D distribution and the predicted one (obtained as the sum of the forward-folded spectra for H, He and the heavy nuclei) at the first iteration step. The iteration procedure stops when the weighted mean squared error is found. The intermediate results were smoothed using the 353HQ-twice routine \cite{smoothing} as implemented in the ROOT package \cite{root}.  The unfolded results are presented just for the interval $\log_{10}(E/\gev) = [4.0, 5.4]$ because the systematic errors increase faster at higher energies and, at lower energies, the effective area decreases by a factor $ < 0.44$.  \vspace{-0.8pc}


\section{Results}

\begin{figure}[!t]
     \centering
     \footnotesize
     \includegraphics[width=75mm]{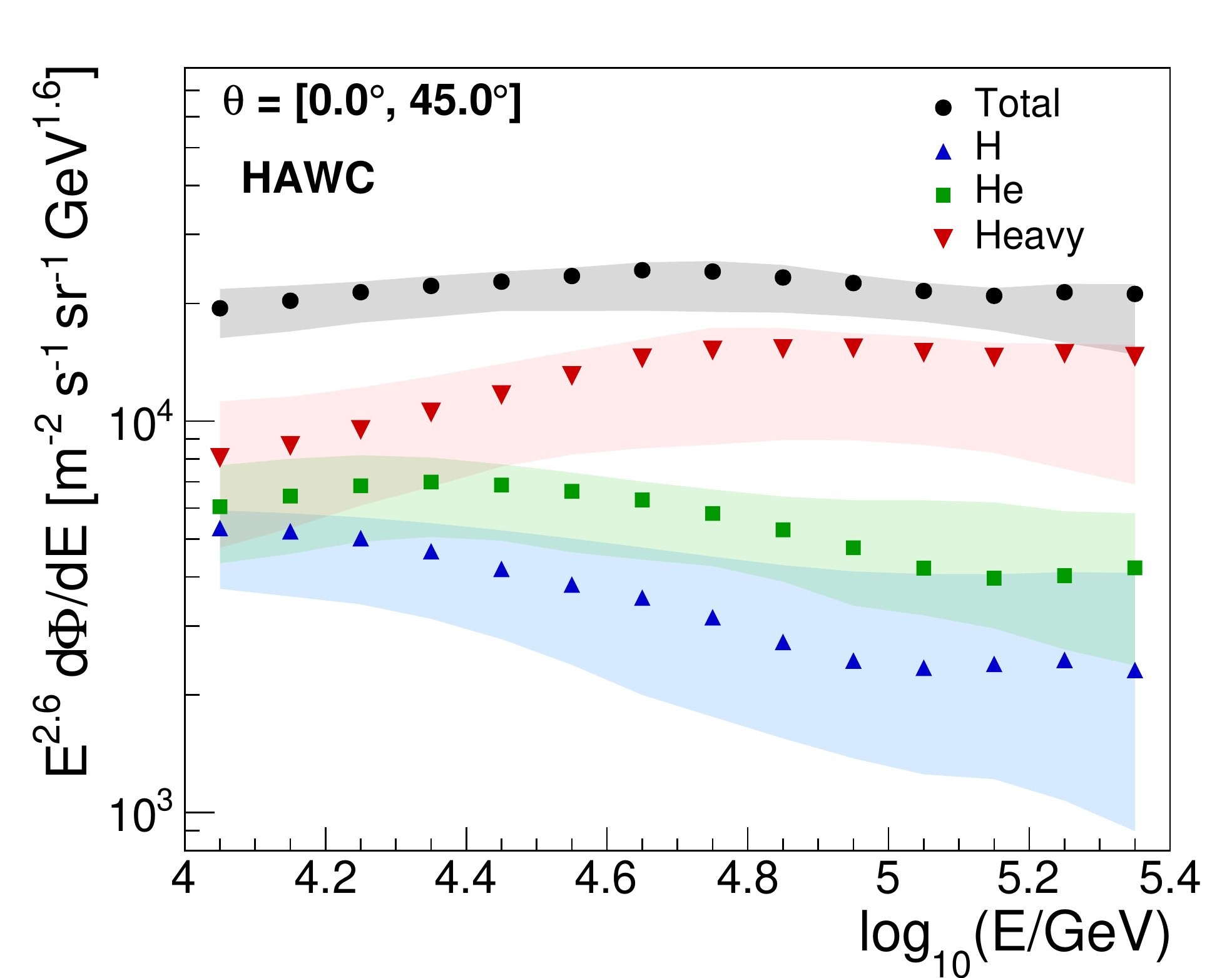}
     \includegraphics[width=75mm]{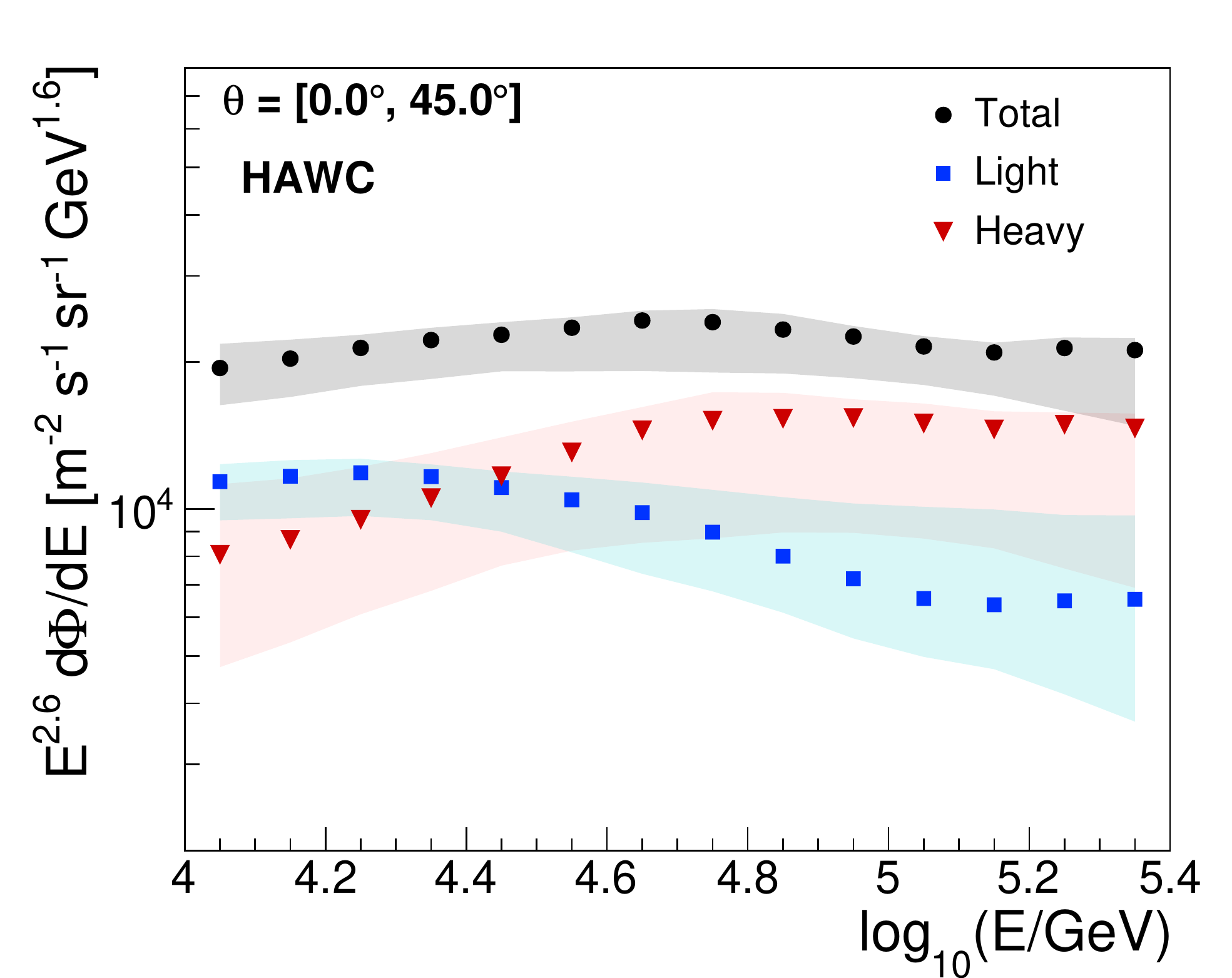} 
     \caption{ 
     The unfolded energy spectra of H (blue upward triangles), He (green squares) and heavy ($Z > 2$) cosmic ray nuclei (red  downward triangles) from HAWC data are shown in the plot on the \textit{left}. The corresponding spectrum for light nuclei (H+He, blue squares) is presented in the figure on the \textit{right} and is compared with the spectrum for heavy  cosmic ray primaries  (red  downward triangles). The all-particle energy spectrum (sum of the unfolded results, dark circles) is shown in both panels. Statistical errors ($< 0.05 \%$) are displayed with vertical error bars and are smaller than the size of the data points. Systematic uncertainties ($< 78 \%$) are shown as error bands. 
    }
  	\label{joinspectra}
  	\vspace{-1pc}
\end{figure} 

 The unfolded energy spectra for proton, helium and heavy mass groups of cosmic rays are shown in fig.~\ref{joinspectra}, left. On the other hand, the sum of the spectra for H and He is compared with the result for the group of heavy elements in fig.~\ref{joinspectra}, right. In these plots, the total intensity, calculated as the sum of the individual elemental spectra is also displayed. The unfolded results are displayed with their corresponding statistical and systematic errors. Statistical errors are due to the limited size of the data sample. Systematic uncertainties include the effect from the statistics of the MC data set, the experimental uncertainties in the parameters of the PMTs \cite{Hawccrab19},  the hadronic interaction model (for which a small MC data set was generated with EPOS-LHC \cite{eposlhc}), the unfolding procedure (bias produced by the method, variations associated to the changes in the seed and the usage of a different unfolding method, in particular, the reduced cross entropy technique), uncertainties in the effective area, the shower age and the composition model used for the calculations (here, we used the Polygonato \cite{poli} and the GSF \cite{gsf} models, and two additional ones calibrated with direct data from ATIC-02 \cite{atic07} and JACEE \cite{jacee}, respectively). The systematic errors are dominated by experimental effects ($< 55 \%$), the shower age ($< 20 \%$) and the hadronic interaction model ($< 30 \%$).  It is important to point out that these systematic errors do not have a drastic effect on the shapes of the spectra, but on the relative light/heavy cosmic ray abundance. 
 
 \begin{figure}[!ht]
     \centering
     \footnotesize
     \includegraphics[width=75mm]{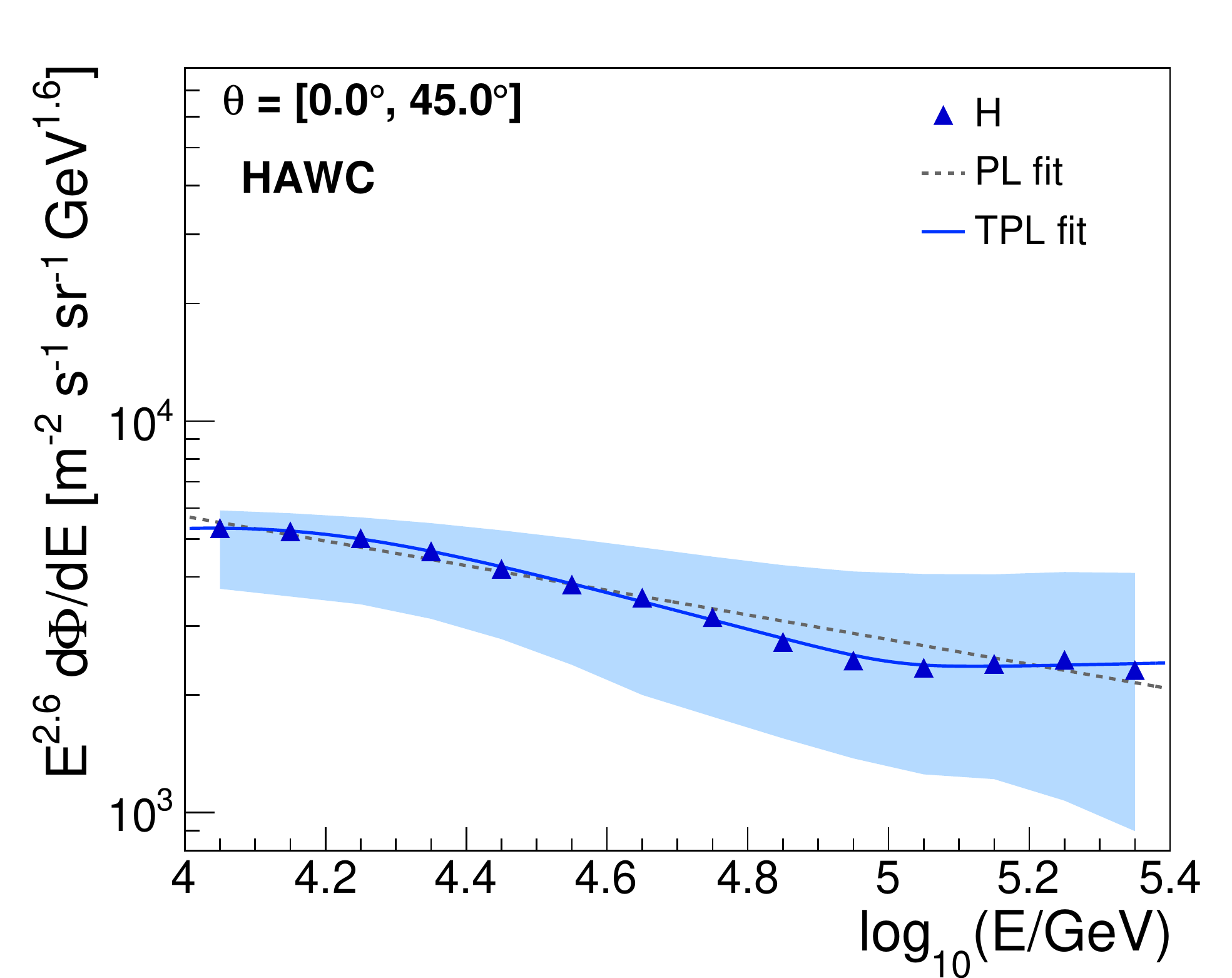}
     \includegraphics[width=75mm]{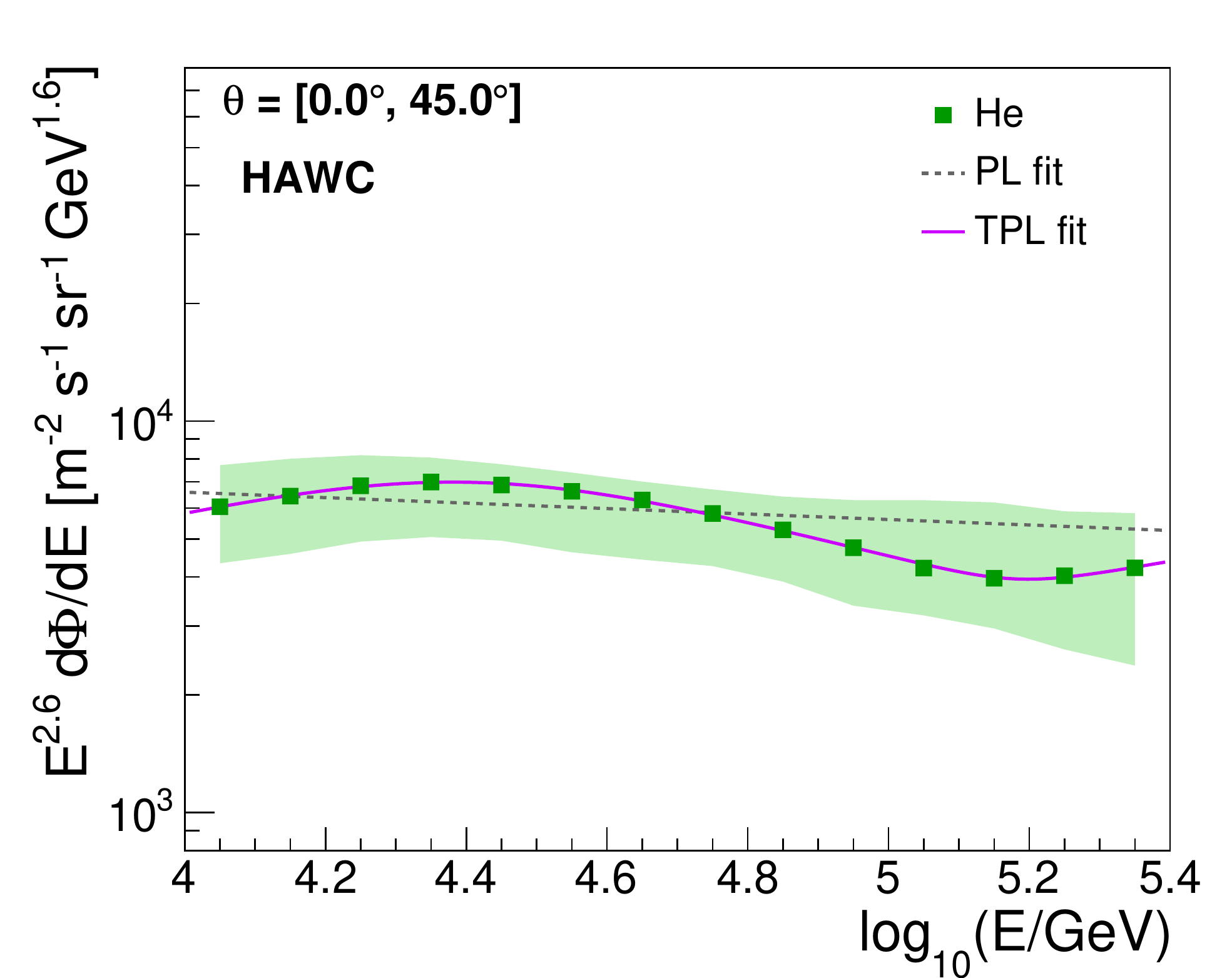} 
     \includegraphics[width=75mm]{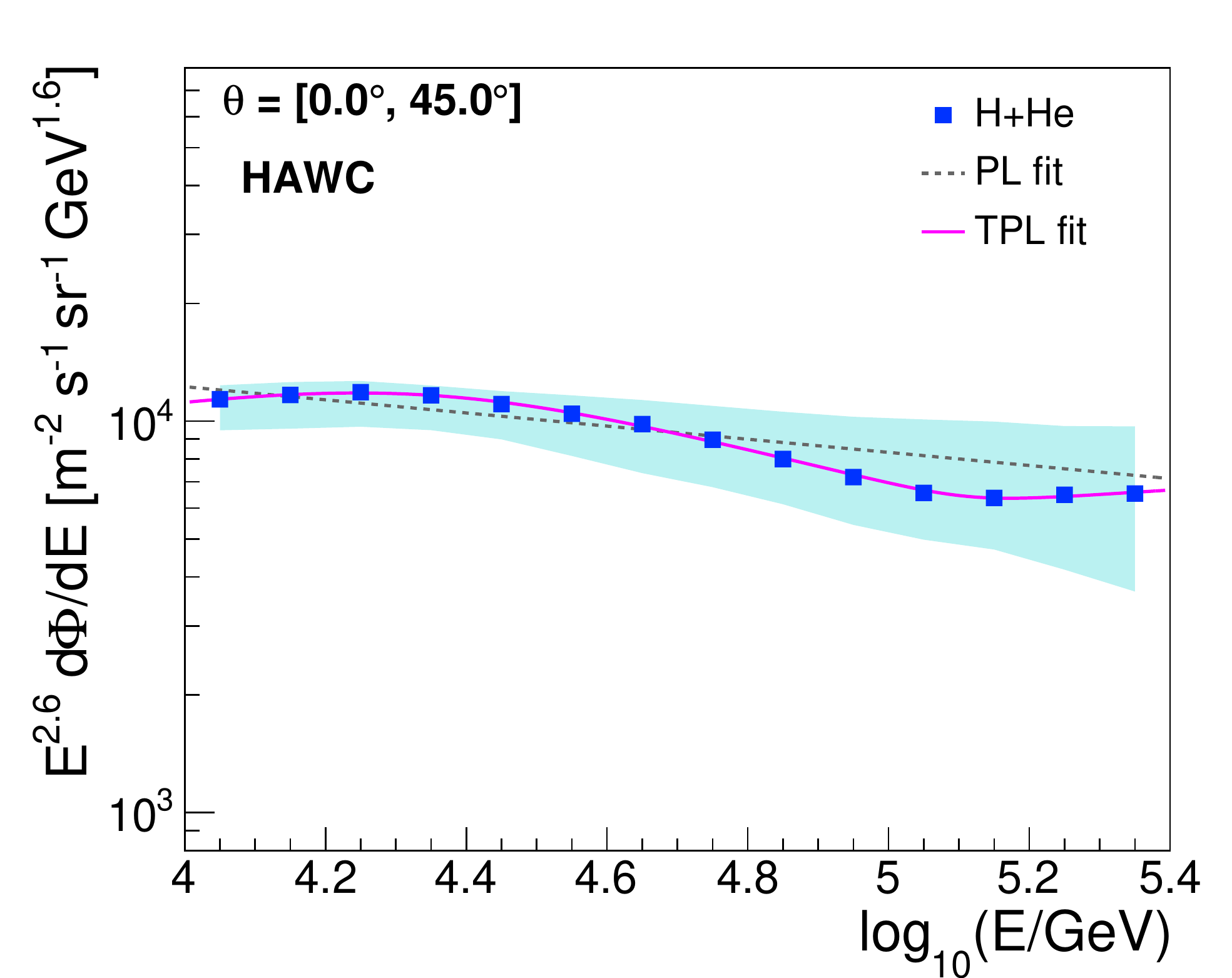}
     \includegraphics[width=75mm]{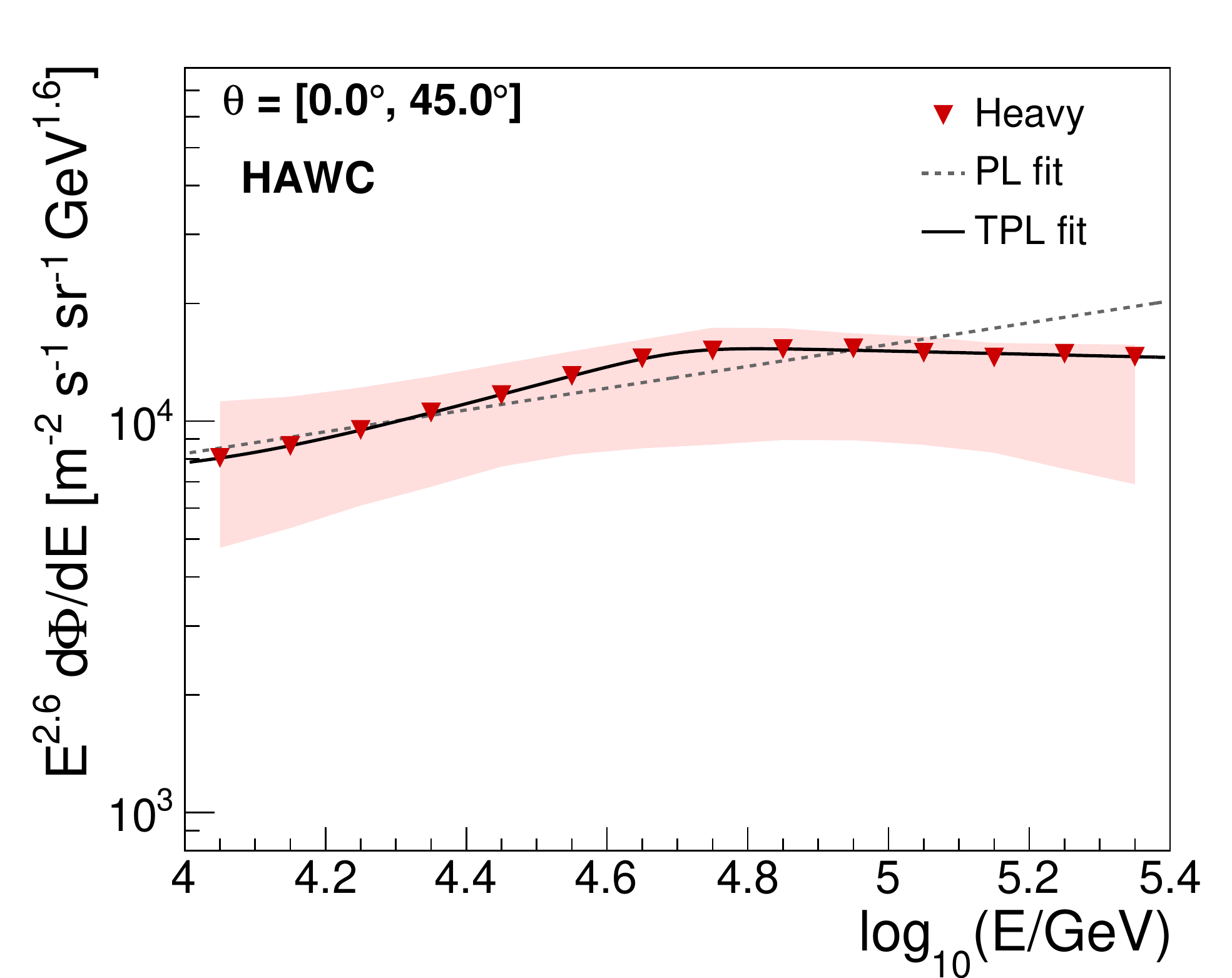} 
     \caption{Fits to the unfolded energy spectra. PL represents the fit with a single power law and TPL, the fit with eq.~(\ref{eq2}).
    }
  	\label{fitspectra}
  	\vspace{-1pc}
\end{figure}

 The spectra presented in fig.~\ref{joinspectra} do not follow a power-law function. HAWC data show the presence of fine structure between $10 \, \tev$ and $251 \, \tev$, in particular, individual softenings at tens of $\tev$ and possible  hardenings in the elemental spectra of protons and helium above $100 \, \tev$. We have fit the elemental spectra with the expression
  \begin{equation}
     \Phi(E) = \Phi_0 \, E^{\gamma_0} \, \left[  1 + \left( \frac{E}{E_0}\right)^{\varepsilon_0}\right]^{(\gamma_1 -\gamma_0)/\varepsilon_0} \, 
     \left[  1 + \left( \frac{E}{E_1}\right)^{\varepsilon_1}\right]^{(\gamma_2 -\gamma_1)/\varepsilon_1},
     \label{eq2}
 \end{equation}
 which is a power-law type function with  two breaks, $E_0$ and $E_1$, at low and high energies, respectively. Here,  $\varepsilon_i$, with $i = 0, 1$, is the smoothing of the $i$ feature, while $\gamma_i$ and $\gamma_{i+1}$ are the spectral indexes of the function before and after the corresponding break. The fitted curves are shown in fig.~\ref{fitspectra} in comparison with the results of a fit with a power-law function. From the curve fitting with eq.~(\ref{eq2}), we found that the cuts observed in the the spectra of H, He and heavy nuclei are located at energies $(14.06 \pm 0.02 (\mbox{stat.}) ^{+2.2}_{-0.4} (\mbox{sys.})) \, \tev$, $(25.30 \pm 0.01 (\mbox{stat.}) ^{+1.1}_{-0.8} (\mbox{sys.})) \, \tev$, $(50.72 \pm 0.01 (\mbox{stat.}) \pm 1 (\mbox{sys.})) \, \tev$, respectively, while for the light mass group spectrum, we observe the softening at $(22.11 \pm 0.02 (\mbox{stat.}) ^{+2.0}_{-0.5} (\mbox{sys.})) \, \tev$. According to these values, the position of the knee-like features increase for heavier nuclei. On the other hand, the fits show that the possible hardenings in the spectra of protons and helium nuclei happens at around $(102.8 \pm 0.1 (\mbox{stat.}) ^{+1}_{-4} (\mbox{sys.})) \, \tev$ and  $(152.2 \pm 0.1 (\mbox{stat.}) ^{+11}_{-9} (\mbox{sys.})) \, \tev$, respectively.  The recovery for the spectra of H+He is observed at  $(130.0 \pm 0.1 (\mbox{stat.}) ^{+3}_{-7} (\mbox{sys.})) \, \tev$. In this regard, a statistical analysis using the test statistics $\Delta \chi^2$  shows that the power-law with two breaks is favored by the data  with  more than $5 \sigma$ of significance over the scenario with only one break in case of the spectra for H, He and H+He.

 We can also observe from fig.~\ref{joinspectra}, that the ratio $\Phi_{H}/\Phi_{He}$ is smaller than one in the analyzed energy interval. Besides, we also see that the composition of cosmic rays becomes heavier at high energies. Finally, we can also notice that the bump in the the all-particle energy spectrum between $10 \, \tev$ and $100 \, \tev$ (previously reported by HAWC in \cite{Hawc17}) is due to the superposition of the individual softenings in the spectra of H, He and heavy nuclei. 

\begin{figure}[!t]
     \centering
     \footnotesize
     \includegraphics[width=75mm]{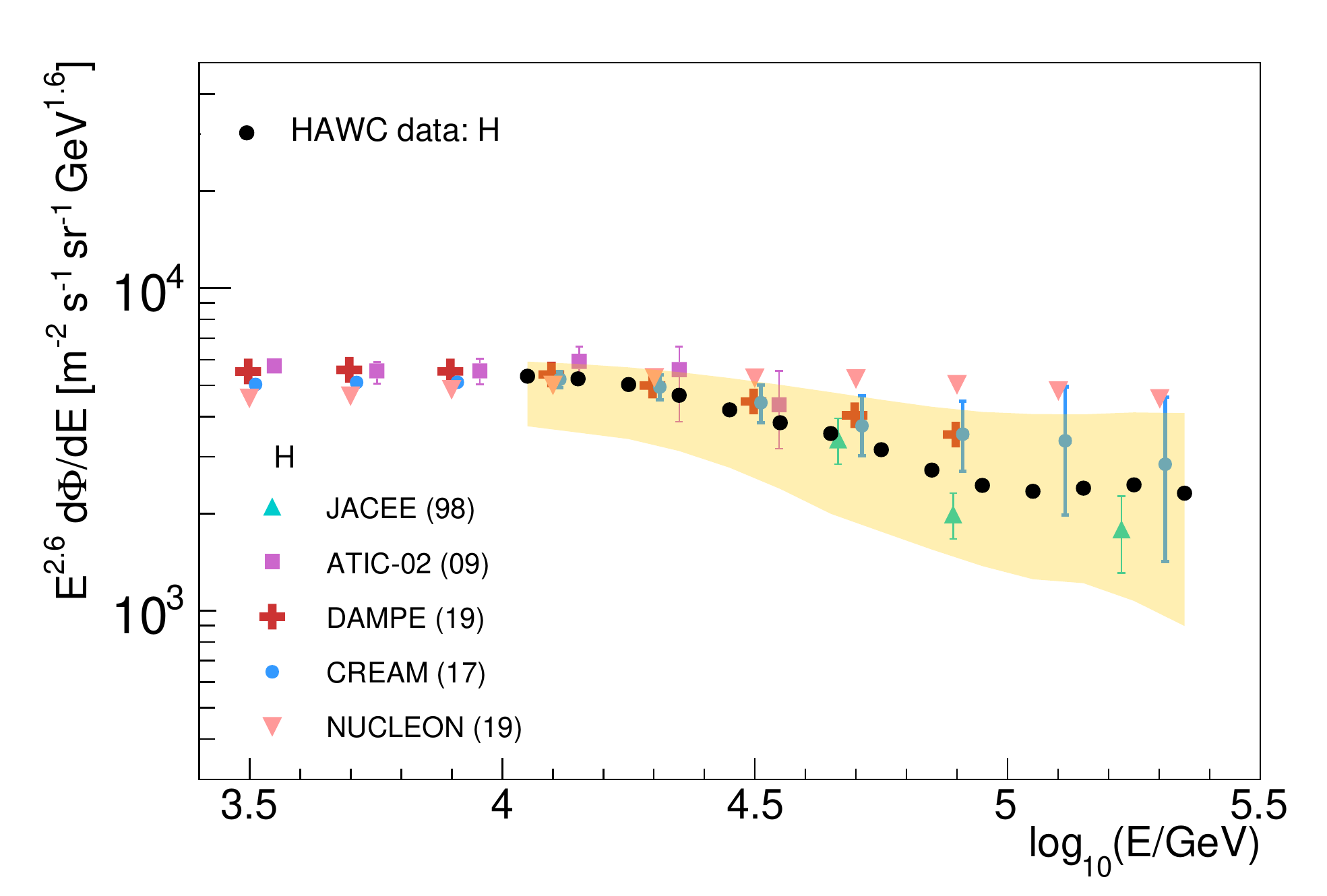}
     \includegraphics[width=75mm]{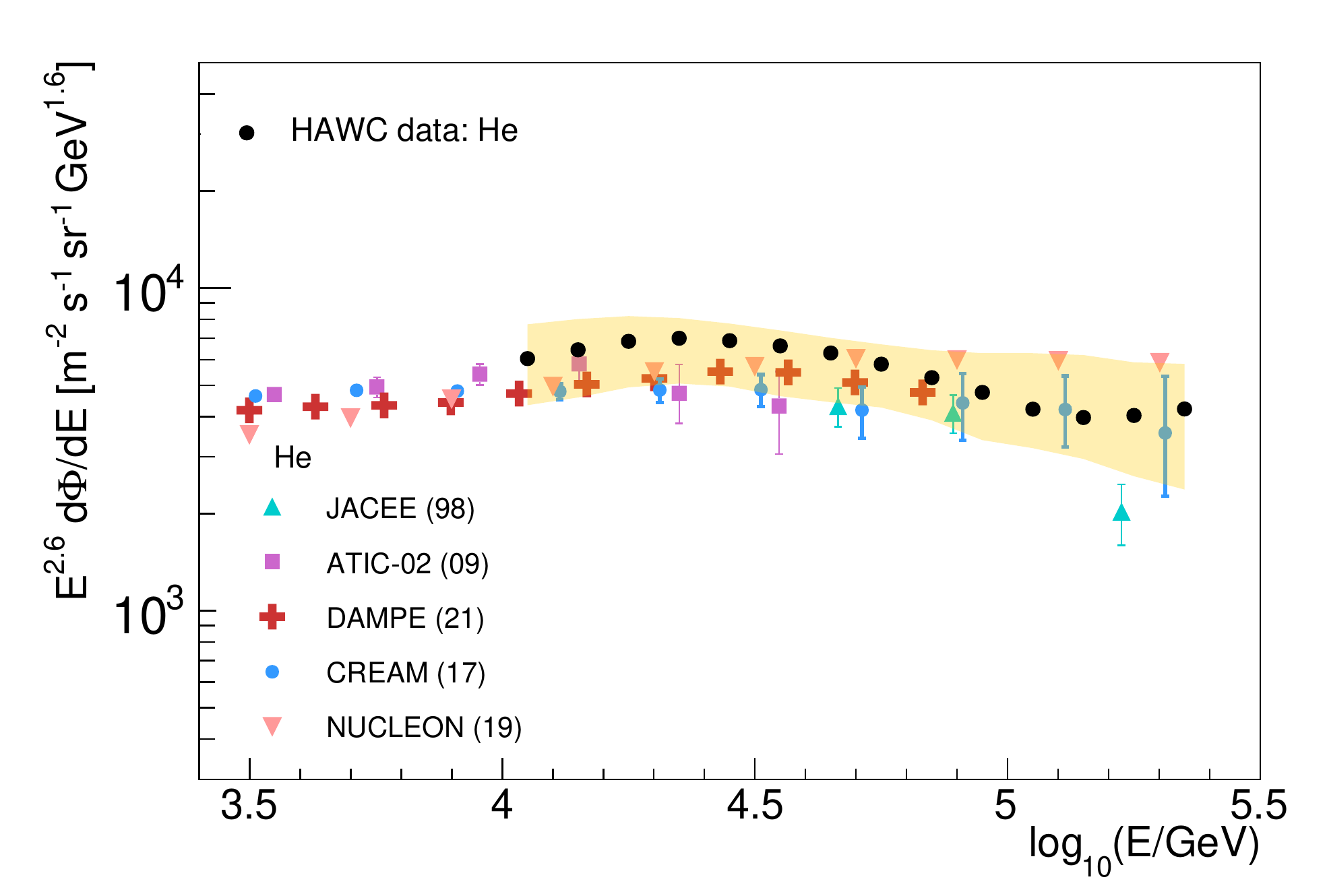} 
     \includegraphics[width=75mm]{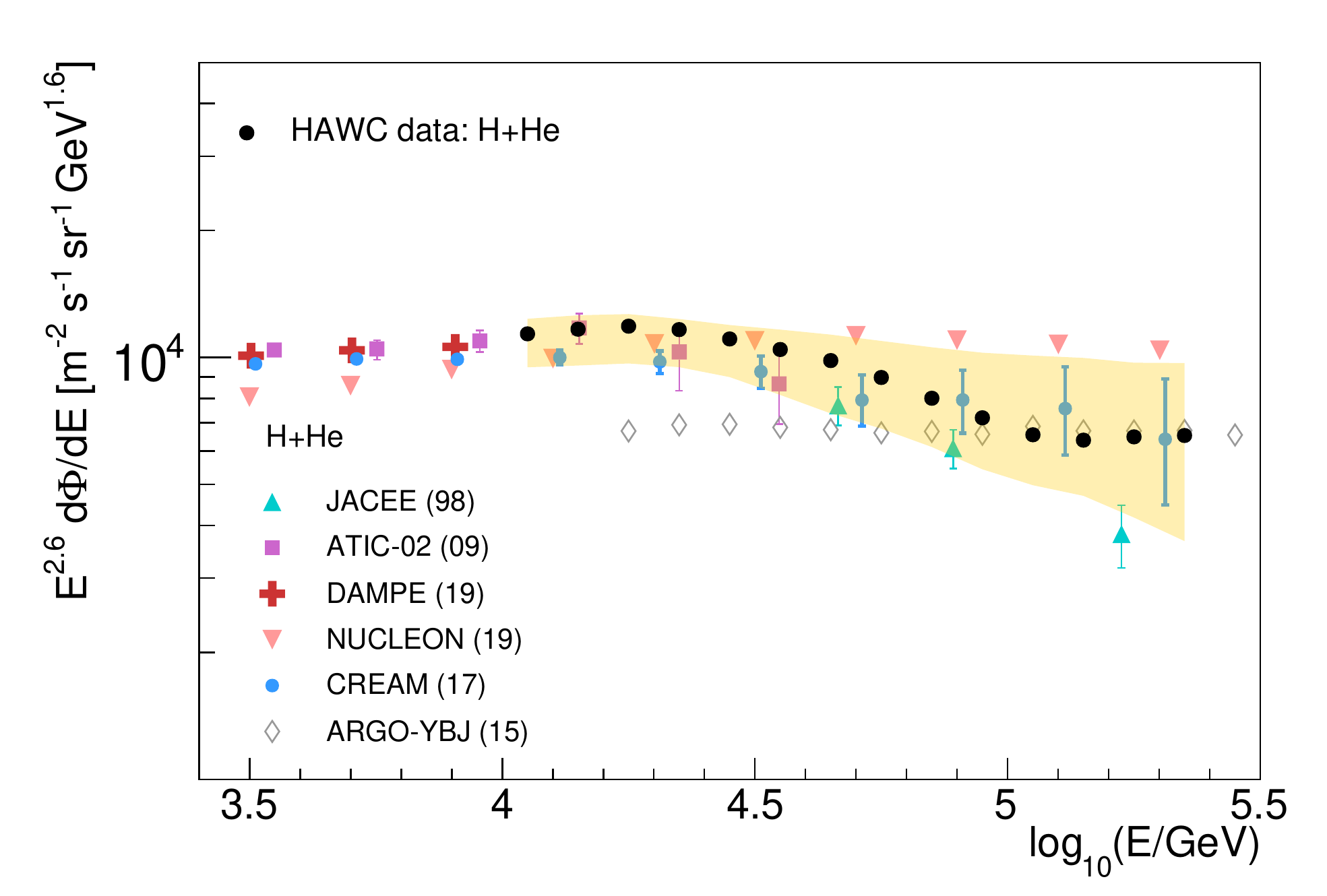}
     \includegraphics[width=75mm]{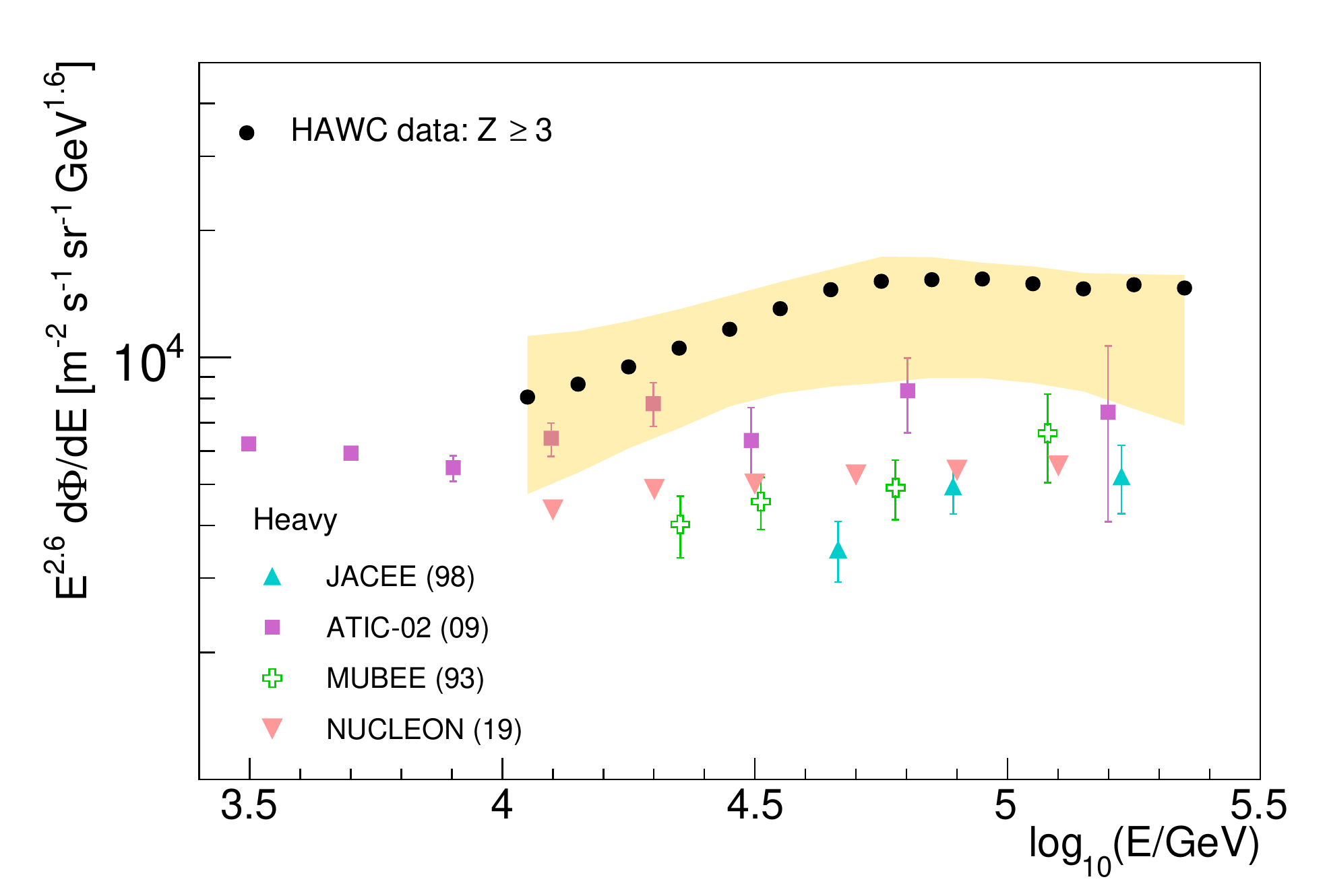} 
     \caption{Comparisons of HAWC results with data from different experiments: JACEE \cite{jacee}, ATIC-02 \cite{atic09}, CREAM \cite{cream17},  DAMPE \cite{dampe19, dampe21}, NUCLEON \cite{nucleon19}, ARGO-YBJ \cite{Argo15} and MUBEE \cite{mubee}.}
  	\label{expcomparisons}
  	\vspace{-1pc}
\end{figure}

\begin{figure}[!t]
     \centering
     \footnotesize
     \includegraphics[width=75mm]{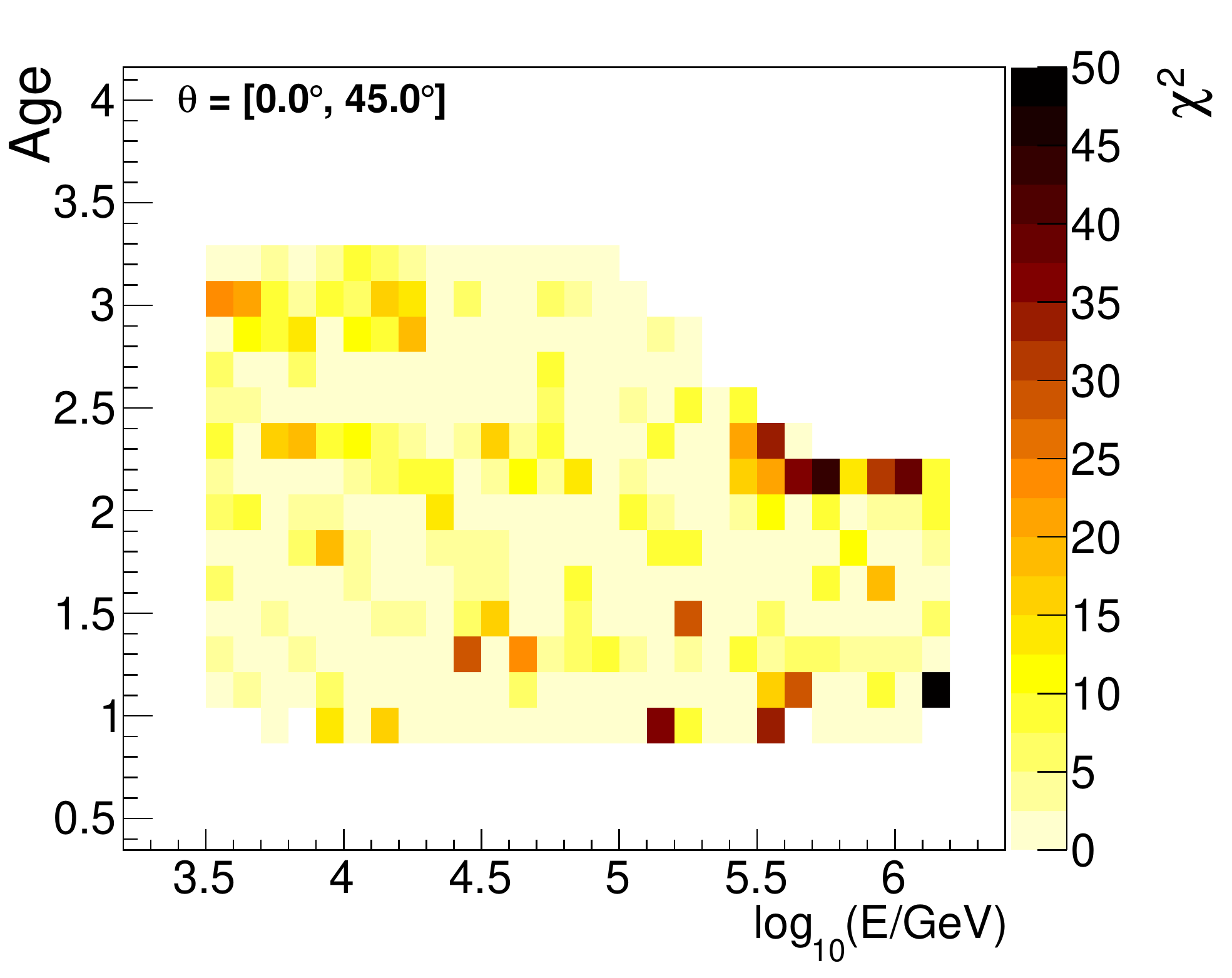}
     \includegraphics[width=75mm]{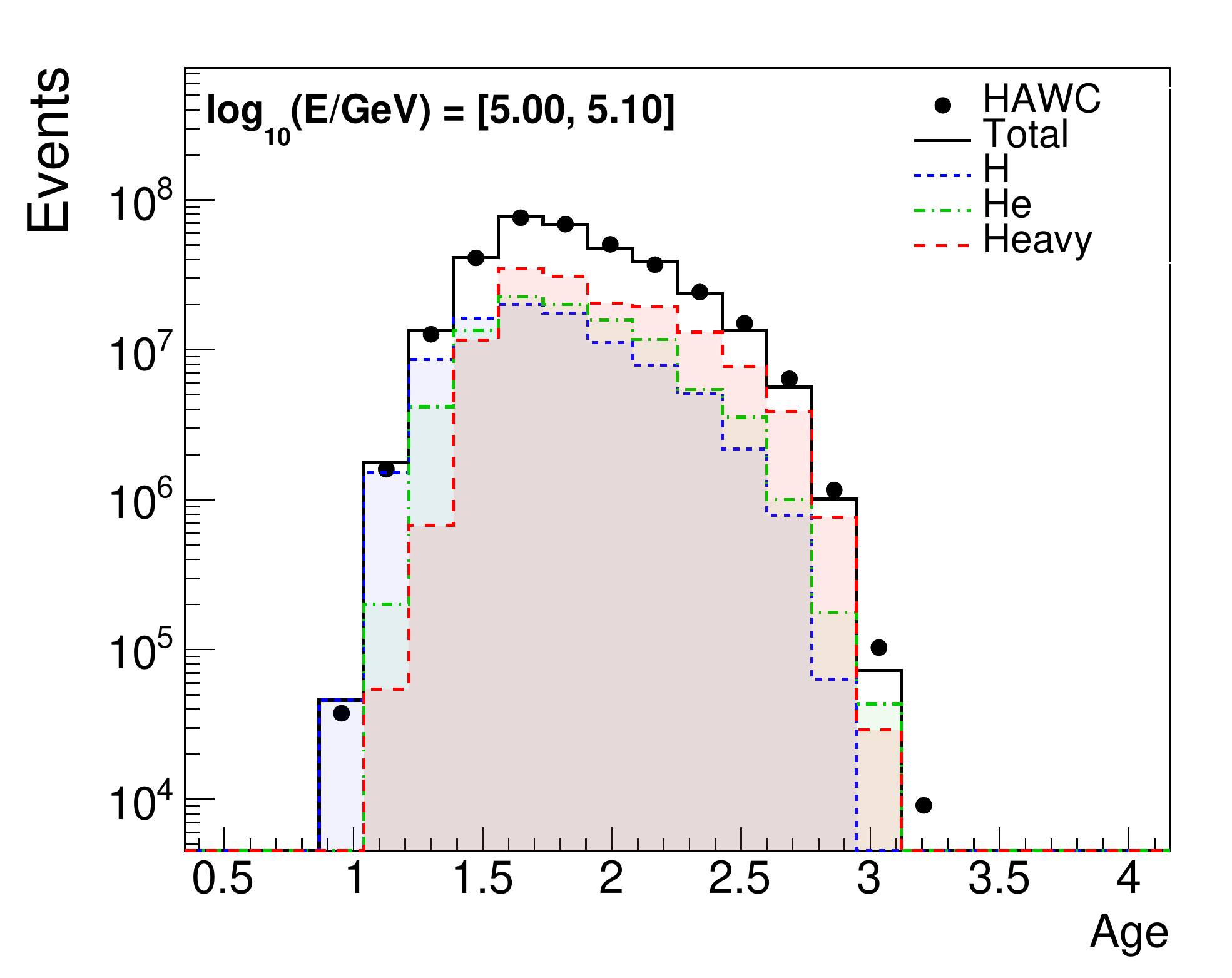} 
     \caption{\textit{Left:} $\chi^2$ per bin in the space of shower age vs reconstructed energy after comparing data vs forward-folded distributions of the solutions. \textit{Right:} Comparison between the forward-folded results (lines) and the data on the shower age (data points) for the energy interval $\log_{10}(E/\gev) = [5.0, 5.1]$.
    }
  	\label{chi2comparison}
  	\vspace{-1pc}
\end{figure}

 The measured spectra are compared in the four panels of fig.~\ref{expcomparisons} with the results of other experiments. We observe that the spectra for protons and He is in good agreement with the direct detection data from JACEE \cite{jacee}, ATIC-02 \cite{atic09}, CREAM \cite{cream17} and DAMPE \cite{dampe19, dampe21} within systematic errors. HAWC results even confirm the softenings in the spectra of protons and helium first hinted by ATIC-02 and CREAM and recently confirmed by DAMPE \cite{dampe19, dampe21}. NUCLEON \cite{nucleon19} results for protons at high energies seems to be above HAWC results, however for He nuclei are in agreement with our measurements.  In the panel for the spectrum of H+He, we also show results from the EAS experiment ARGO-YBJ \cite{Argo15}. HAWC data for H+He are in good agreement with the direct measurements from JACEE \cite{jacee}, ATIC-02 \cite{atic09} and CREAM \cite{cream17}. However, HAWC results agree with ARGO-YBJ only above $300 \, \tev$. We also confirmed the knee-like feature reported in \cite{Hawc19} by HAWC in the $\tev$ region of the spectrum for the light cosmic ray primaries. The spectrum for the heavy mass group ($Z > 2$) obtained with HAWC is in agreement with ATIC-02 within systematic uncertainties, but it is above the data from NUCLEON, MUBEE \cite{mubee} and JACEE. This shift upwards of the spectrum for the heavy component of cosmic rays seems to be mainly related with contamination from light nuclei as a result of a systematic on the shower age. Finally, in fig.~\ref{chi2comparison}, left, we show the $\chi^2$ distribution, in the $s$ vs $\log_{10}(E)$ space, obtained from a comparison between the data and the sum of the forward-folded results. In each bin the squared deviations have been divided by the corresponding sum in quadrature of the statistical errors of both 2D histograms. We can see that, in general, the solutions provide a reasonable description of the data, except for old EAS at low and high energies. A more detailed comparison is shown in  fig.~\ref{chi2comparison}, right, where we see the measured age distributions and the corresponding forward-folded solutions for one $\log_{10}(E)$ energy bin. The total $\chi^2$ per number of degrees of freedom is $5.02$.  \vspace{-1pc}

\section{Conclusions}

 We have estimated the energy spectrum of cosmic rays with HAWC for three elemental mass groups: H, He and heavy nuclei ($Z > 2$) between $10 \, \tev$ and $251 \, \tev$. HAWC results show that the spectra of these mass groups have fine structures, in particular, individual softenings, whose energy positions increase with the primary mass.  The observation of softenings in the spectra of H and He with HAWC, close to $(14.06 \pm 0.02 (\mbox{stat.}) ^{+2.2}_{-0.4} (\mbox{sys.})) \, \tev$ and $(25.30 \pm 0.01 (\mbox{stat.}) ^{+1.1}_{-0.8} (\mbox{sys.})) \, \tev$, respectively, confirms the recent detections with the DAMPE satellite of similar features in the intensity of protons \cite{dampe19}, at around  $(13.6^{+4.1}_{-4.8}) \, \tev$,  and helium \cite{dampe21}, at approximately $(34.4^{+6.7}_{-9.8}) \, \tev$. This is the first time that high-statistics data on cosmic ray composition from direct and EAS experiments are compared. The good agreement between the results from both techniques confirms the potential of high-altitude EAS observatories like HAWC for the research of $\tev$ cosmic rays. In addition, there is a new feature in the spectrum of the heavy component of cosmic rays in the TeV region and indications in the HAWC data in favor of possible hardenings in the intensities of protons and helium close to $100 \, \tev$.  On the other hand,  the relative abundance of $\Phi_{H+He}(E)/\Phi_{Z>2}$ is observed to decrease from $10 \, \tev$ to $100 \, \tev$ mainly due to the cut in the spectrum of the light component of cosmic rays. Interestingly, the bump previously observed in the all-particle energy spectrum of cosmic rays by HAWC close to $(45.7 \pm 1.1)\,  \tev$ \cite{Hawc17} is produced by the superposition of the cuts in the elemental spectra of H, He and heavy cosmic ray primaries. We also observed that the superposition of the knee-like features in the H and He spectra generates the $\tev$ softening reported in \cite{Hawc19} by HAWC for the spectrum of the light cosmic ray species. Finally, we notice from HAWC data that the softenings for H and He nuclei seem to occur at the same rigidity value as pointed out in \cite{nucleon19} by the NUCLEON collaboration. In particular, we observe in our results that the ratio between the energy cuts for He and H nuclei is $E_{cut, He}/E_{cut, H} \sim 1.8^{+0.3}_{-0.1}$, which is around the expected value of $2$ in the rigidity dependent scenario.\\ \vspace{-0.5pc}


  \noindent \textbf{Acknowledgments}. {\scriptsize
  We acknowledge the support from: the US National Science Foundation (NSF); the US Department of Energy Office of High-Energy Physics; the Laboratory Directed Research and Development (LDRD) program of Los Alamos National Laboratory; Consejo Nacional de Ciencia y Tecnolog\'ia (CONACyT), M\'exico, grants 271051, 232656, 260378, 179588, 254964, 258865, 243290, 132197, A1-S-46288, A1-S-22784, c\'atedras 873, 1563, 341, 323, Red HAWC, M\'exico; DGAPA-UNAM grants IG101320, IN111716-3, IN111419, IA102019, IN110621, IN110521; VIEP-BUAP; PIFI 2012, 2013, PROFOCIE 2014, 2015; the University of Wisconsin Alumni Research Foundation; the Institute of Geophysics, Planetary Physics, and Signatures at Los Alamos National Laboratory; Polish Science Centre grant, DEC-2017/27/B/ST9/02272; Coordinaci\'on de la Investigaci\'on Cient\'ifica de la Universidad Michoacana; Royal Society - Newton Advanced Fellowship 180385; Generalitat Valenciana, grant CIDEGENT/2018/034; Chulalongkorn University’s CUniverse (CUAASC) grant; Coordinaci\'on General Acad\'emica e Innovaci\'on (CGAI-UdeG), PRODEP-SEP UDG-CA-499; Institute of Cosmic Ray Research (ICRR), University of Tokyo, H.F. acknowledges support by NASA under award number 80GSFC21M0002. We also acknowledge the significant contributions over many years of Stefan Westerhoff, Gaurang Yodh and Arnulfo Zepeda Dominguez, all deceased members of the HAWC collaboration. Thanks to Scott Delay, Luciano D\'iaz and Eduardo Murrieta for technical support.
  }
  

%
\clearpage
\section*{Full Authors List: HAWC Collaboration}

\scriptsize
\noindent
A.U. Abeysekara$^{48}$,
A. Albert$^{21}$,
R. Alfaro$^{14}$,
C. Alvarez$^{41}$,
J.D. Álvarez$^{40}$,
J.R. Angeles Camacho$^{14}$,
J.C. Arteaga-Velázquez$^{40}$,
K. P. Arunbabu$^{17}$,
D. Avila Rojas$^{14}$,
H.A. Ayala Solares$^{28}$
% ...

\noindent
% ...
$^{14}$Instituto de Física, Universidad Nacional Autónoma de México, Ciudad de México, México,
% ...
$^{17}$Instituto de Geofísica, Universidad Nacional Autónoma de México, Ciudad de México, México,
% ...
$^{21}$Physics Division, Los Alamos National Laboratory, Los Alamos, NM, USA,
% ...
$^{28}$Department of Physics, Pennsylvania State University, University Park, PA, USA,
% ...
$^{40}$Instituto de Física y Matemáticas, Universidad Michoacana de San Nicolás de Hidalgo, Morelia, Michoacán, México,
$^{41}$FCFM-MCTP, Universidad Autónoma de Chiapas, Tuxtla Gutiérrez, Chiapas, México,
% ...
$^{48}$Department of Physics and Astronomy, University of Utah, Salt Lake City, UT, USA,
% ...

\begin{thebibliography}{99}
\scriptsize
 \bibitem{Hawc17} R. Alfaro et al., HAWC Collaboration, Phys. Rev. D 96 (2017) 122001.
 \vspace{-0.8pc}
 
 \bibitem{Hawc17a} A. U. Abeysekara et al., HAWC Collaboration, Astrophys. J. 843 (2017) 39.
 \vspace{-0.8pc}

 \bibitem{qgsjetii4} S. Ostapchenko, Phys. Rev. D 83 (2011) 014018.
 \vspace{-0.8pc}

 \bibitem{Hawccrab19} A. U. Abeysekara et al., HAWC Collaboration, Astrophys. J. 881 (2019) 134.
 \vspace{-0.8pc}

 \bibitem{Heck:1998vt} D. Heck et al., CORSIKA: A Monte Carlo Code to Simulate Extensive Air Showers, FZK Berichte 6019, Karlsruhe, Germany, 1998.
 \vspace{-0.8pc}

 \bibitem{Fluka} A. Ferrari et al., FLUKA: a multi-particle transport code, Report INFN/TC\_05/11, SLAC-R-773, CERN-2005-10, 2005.
 \vspace{-0.8pc}

 \bibitem{pamela} O. Adriani et al., PAMELA Collaboration, Science 332 (2011) 69.
 \vspace{-0.8pc}

 \bibitem{ams14} M. Aguilar et al., AMS Collaboration, Phys. Rev. Lett. 114 (2015) 171103.
 \vspace{-0.8pc}

 \bibitem{ams15} M. Aguilar et al., AMS Collaboration, Phys. Rev. Lett. 115 (2015) 211101.
 \vspace{-0.8pc}
 
 \bibitem{cream09} H. S. Ahn et al., CREAM Collaboration, Astrophys. J. 707 (2009) 593.
 \vspace{-0.8pc}

 \bibitem{cream11} Y. S. Yoon et al., CREAM Collaboration, Astrophys. J. 728 (2011) 122.
 \vspace{-0.8pc}
 
 \bibitem{Gold64} R. Gold, An iterative unfoding method for response matrices, Report ANL-6984,
 Argonne National Laboratory, USA,1964.
 \vspace{-0.8pc}

  \bibitem{Ulrich01} H. Ulrich et al., KASCADE Collaboration, Proc. of the 27th ICRC (Hamburg, Germany), 97, 2001.
 \vspace{-0.8pc}
  
  \bibitem{Crossentrop} M. Schmelling, NIMA 340 (1994) 400.
 \vspace{-0.8pc}
  
  \bibitem{smoothing} J. Friedman, Proceedings of the 1974 Cern School of computing (Norway), 1974, 271.
 \vspace{-0.8pc}
  
  
  \bibitem{root} R. Brun and F. Rademakers, NIMA 389 (1997) 81.
 \vspace{-0.8pc}
 
  \bibitem{eposlhc} T. Pierog et al., Phys. Rev. C 92 (2015) 034906.
 \vspace{-0.8pc}

 \bibitem{poli} J. R. Hoerandel, Astropart. Phys. 19 (2003) 193.
 \vspace{-0.8pc}
 
 \bibitem{gsf} H. P. Dembinski et al., PoS(ICRC2017) 533.
 \vspace{-0.8pc}


  \bibitem{atic07} A. D. Panov et al., ATIC-2 Collaboration, Bull. Russ. Acad. Sci. Phys. 71 (2007)  494.
 \vspace{-0.8pc}


 \bibitem{jacee} Y. Takahashi et al., JACEE Collaboration, Nucl. Phys. B (Proc. Suppl.) 
 60 (1998) 83.
 \vspace{-0.8pc}


 \bibitem{atic09} A. D. Panov et al., ATIC-2 Collaboration, Bull. Russ. Acad. Sci. Phys. 73, No. 5 (2009) 564.
 \vspace{-0.8pc}
 
  \bibitem{cream17} Y. S. Yoon et al., CREAM Collaboration, Astrophys. J. 839 (2017) 5.
 \vspace{-0.8pc}


 \bibitem{dampe19} Q. An et al., DAMPE Collaboration, Science Advances 5, No. 9 (2019) eaax3793.
 \vspace{-0.8pc}

 \bibitem{dampe21} F. Alemanno et al., DAMPE Collaboration, PRL 126 (2021)  201102.
 \vspace{-0.8pc}
 
 
  \bibitem{nucleon19} E. V. Atkin et al., NUCLEON Collaboration, Astron. Rep. 63 (2019) 66.
 \vspace{-0.8pc}

  \bibitem{Argo15} B. Bartoli et al., ARGO-YBJ Collaboration, PRD 92 (2015) 092005.
 \vspace{-0.8pc}

 \bibitem{mubee} V. I. Zatsepin et al., MUBEE Collaboration, Proc. of the 23rd ICRC (Calgary, Canada), Vol. 2, 1993, No. 13.
 \vspace{-0.8pc}
  
 \bibitem{Hawc19} J. C. Arteaga-Vel\'azquez et al., HAWC Collaboration, PoS(ICRC2019) 176.
\end{thebibliography}
\end{document}